\documentclass[apjl]{emulateapj-rtx4}
\def\psr{{\rm PSR J0437$-$4715}}

\def\psrhs{{\rm PSR B1737+13}}

\def\Curtin{$^{1}$}
\def\CAASTRO{$^{2}$}

\newcommand{\be}{\begin{eqnarray}}
\newcommand{\ee}{\end{eqnarray}}

\renewcommand{\vec}[1]{\mathbf{#1}}

\newcommand{\fnu}{\mbox{${f_{\nu} }$\,}}
\newcommand{\ft}{\mbox{${f _{\rm t} }$\,}}
\newcommand{\ftsq}{\mbox{${f ^2 _{\rm t} }$\,}}
\newcommand{\fti}{\mbox{${f _{\rm t,i} }$\,}}
\newcommand{\ftisq}{\mbox{${f ^2 _{\rm t,i} }$\,}}

\def\la{\mbox{\raisebox{-0.1ex}{$\scriptscriptstyle \stackrel{<}{\sim}$\,}}}
\def\ga{\mbox{\raisebox{-0.1ex}{$\scriptscriptstyle \stackrel{>}{\sim}$\,}}}
\newcommand{\nuobs}{\mbox{${\rm \nu _{obs} }$\,}}
\newcommand{\Viss}{\mbox{${V _{\rm iss} }$\,}}

\newcommand{\Vmu}{\mbox{${V _{\rm \mu} }$\,}}

\newcommand{\Vearth}{\mbox{${V _{\rm earth,\perp} }$\,}}
\newcommand{\Vscreen}{\mbox{${V _{\rm scr} }$\,}}
\newcommand{\vecVeff}{\mbox{${\vec{V} _{\rm eff} }$\,}}
\newcommand{\Veff}{\mbox{${V _{\rm eff} }$\,}}

\newcommand{\Vpsrperp}{\mbox{${V _{\rm \mu\bot} }$\,}}
\newcommand{\vecVpsrperp}{\mbox{${\vec{V} _{\rm \mu\bot} }$\,}}

\newcommand{\vecVscreenperp}{\mbox{${\vec{V} _{\rm scr \bot} }$\,}}
\newcommand{\Vobs}{\mbox{${V _{\rm earth} }$\,}}
\newcommand{\Vobsperp}{\mbox{${V _{\rm earth \perp} }$\,}}
\newcommand{\vecVobsperp}{\mbox{${\vec{V} _{\rm earth \perp} }$\,}}

\newcommand{\Veffsq}{\mbox{${V ^2 _{\rm eff } }$\,}}
\newcommand{\Veffperp}{\mbox{${V _{\rm eff \perp} }$\,}}
\newcommand{\vecVeffperp}{\mbox{${\vec{V} _{\rm eff \perp} }$\,}}
\newcommand{\Vbinperp}{\mbox{${V _{\rm bin\perp} }$\,}}
\newcommand{\vecVbinperp}{\mbox{${\vec{V} _{\rm bin\perp} }$\,}}
\newcommand{\Parc}{\mbox{${P _{\rm arc} }$\,}}
\newcommand{\Ntile}{\mbox{${N _{\rm tile} }$\,}}

\newcommand{\etamwa}{${\rm  \eta _{\rm mwa} }$\,}
\newcommand{\etapks}{${\rm  \eta _{\rm pks} }$\,}
\newcommand{\etau}{${\rm  s ^ {3}}$\,}

\newcommand{\taudiff}{\mbox{${\tau _{\rm diff} }$\,}}
\newcommand{\tauref}{\mbox{${\tau _{\rm ref} }$\,}}

\newcommand{\nud}{\mbox{${\nu _{\rm d} }$\,}}
\newcommand{\taud}{\mbox{${\tau _{\rm d} }$\,}}

\newcommand{\dtnu}{\mbox{${d t / d \nu }$\,}}

\newcommand{\driftunit}{${\rm s \ MHz ^ {-1}}$\,}
\newcommand{\cnsqunits}{${\rm m ^{-20/3}}$\,}

\newcommand{\dmu}{${\rm pc \ cm ^ {-3}}$\,}
\newcommand{\velu}{${\rm km \ s ^ {-1}}$\,}
\newcommand{\muu}{${\rm mas \ yr^ {-1}}$\,}

\newcommand{\thetadiff}{\mbox{${{\rm \theta _{diff}}}$\,}}

\newcommand{\thetaone}{\mbox{${{\rm {\theta} _{1}}}$\,}}
\newcommand{\vecthetaone}{\mbox{${{\vec{\theta} _{\rm 1}}}$\,}}
\newcommand{\thetatwo}{\mbox{${{\rm {\theta} _{2}}}$\,}}
\newcommand{\vecthetatwo}{\mbox{${{\vec{\theta} _{\rm 2}}}$\,}}

\newcommand{\Ds}{\mbox{${D _{\rm s}}$\,}}

\newcommand{\Cnsq}{\mbox{${ \overline { C _{\rm n} ^2 } }$\,}}

\shortauthors{Bhat et al.}
\shorttitle{ Scintillation arcs in  \psr\ }
\begin{document}
\title{Scintillation arcs in low-frequency observations of  the timing-array millisecond pulsar \psr}
\medskip
\author{
N. D. R. Bhat\Curtin$^,$\CAASTRO,
S. M. Ord\Curtin$^,$\CAASTRO,
S. E. Tremblay\Curtin$^,$\CAASTRO,
S. J. McSweeney\Curtin$^,$\CAASTRO,
S. J. Tingay\Curtin$^,$\CAASTRO}
\affil{
$^{1}$International Centre for Radio Astronomy Research, Curtin University, Bentley, WA 6102, Australia\\
$^{2}$ARC Centre of Excellence for All-sky Astrophysics (CAASTRO)
}
\medskip
\begin{abstract}
Low-frequency observations of pulsars provide a powerful means for probing the microstructure in the turbulent interstellar medium (ISM). Here we report on high-resolution dynamic spectral analysis of our observations of the timing-array millisecond pulsar \psr\ with the Murchison Widefield Array (MWA), enabled by our recently commissioned tied-array beam processing pipeline for voltage data recorded from the high time resolution mode of the MWA. A secondary spectral analysis reveals faint parabolic arcs, akin to those seen in high-frequency observations of pulsars with the Green Bank and Arecibo telescopes. Data from Parkes observations at a higher frequency of 732\,MHz reveal a similar parabolic feature, with a curvature that scales approximately as the square of the observing wavelength ($\lambda^2$) to the MWA's frequency of 192\,MHz. Our analysis suggests that scattering toward \psr\ predominantly arises from a compact region about 115\,pc from the Earth, which matches well with the expected location of the edge of the Local Bubble that envelopes the local Solar neighbourhood. As well as demonstrating new and improved pulsar science capabilities of the MWA, our analysis underscores the potential of low-frequency pulsar observations for gaining valuable insights into the local ISM and for characterising the ISM toward timing-array pulsars. \\
\end{abstract}

\section{Introduction} \label{s:intro}

Pulsar signals are subjected to a range of delays, distortions and amplitude modulations from dispersive and scattering effects due to the ionised interstellar medium (ISM). 
These propagation effects scale steeply with the observing frequency,  making low frequencies ($\la$400 MHz) less appealing for high-precision timing experiments such as pulsar timing arrays \citep[PTAs;][]{ppta,nanograv,epta}.
However, with  PTA experiments approaching timing precisions of $\sim$0.1-0.8 $\mu$s, there is renewed interest in understanding 
the local ISM and its effects on high-precision timing, which may potentially limit achievable timing precision for most PTA pulsars \citep{nanograv+2015,epta+2015,ppta+2013}. 
While the effects of temporal variations in dispersion measure (DM) have been investigated to a certain extent 
\citep{you+2007,cordes2010,keith+2013,lee+2014,lam+2015,cordes+2015}, there exists only a limited understanding of the impact of scattering on timing precision.\\

PTA  experiments currently rely on millisecond pulsars (MSPs) with low to moderate DMs ($\la$50\,\dmu) in order to minimise ISM effects on timing precision. Recent work that used the Parkes pulsar timing array (PPTA) data to place a limit on the strength of the stochastic gravitational-wave background \citep{ppta+2013} advocates shorter-wavelength (\ga3 GHz) observations to alleviate ISM effects. However, this is not currently feasible for the majority of PTA pulsars, for which $\sim$1-2 GHz remains the most practically viable choice due to sensitivity limitations of existing telescopes and instrumentation. The ISM effects, including multi-path scattering, may still be significant at those frequencies. Since scattering delays ($\taud$) scale steeply with the frequency \citep[$\taud\propto\nu^{-4}$, where $\nu$ is the observing frequency;][]{bhat+2004}, they are more readily measurable in observations with new low-frequency arrays such as the MWA \citep{tingay+2013}, the Long Wavelength Array \citep[LWA;][]{lwa2012} and Low Frequency Array \citep[LOFAR;][]{lofar2013}. Early pulsar observations with these instruments already demonstrate their potential in this direction \citep{bhat+2014,dowell+2013,anne+2014}. \\

Observations of ``scintillation arcs" -- faint, parabolic arc-like features seen in secondary spectral analysis of pulsar  observations have provided new insights into both the micro-structure of the ISM and the interstellar scattering phenomenon \citep{cordes+2006,rickett2007,stinebring2007}. First recognised by \citet{stine+2001} in Arecibo data, detailed studies of these arcs have revealed a variety and richness in their observational manifestations; e.g. forward and reverse arcs, and a chain of arclets \citep{ps2006,hill+2005,brisken+2010}, which  also stimulated a great deal of  theoretical and modelling work \citep{cordes+2006,walker+2004,brisken+2010}. In the context of PTAs,  \citet{hs2008} measured scattering delays from their observations of scintillation arcs in \psrhs\ (DM=48.9\,\dmu); the delays varied from $\sim$0.2 to 2 $\mu$s at a frequency of 1.3 GHz. 

In this paper we present our observations of parabolic scintillation arcs in \psr, a high-priority target for PTAs. Details on processing and analysis are described in \S~2 and \S~3, while in \S 4 and \S 5 we describe the estimation of the arc curvature and the placement of the scattering screen. Our conclusions and future prospects are summarised in \S 6. 

\begin{figure*}[t]
\begin{center}
\epsscale{1.5}
\includegraphics[width=3.18cm,angle=270]{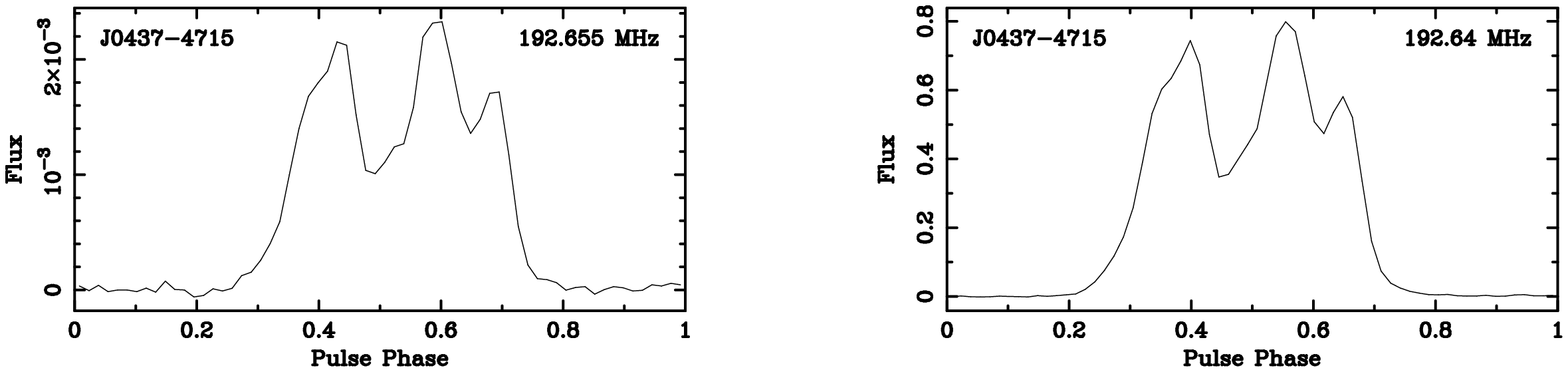}
\vskip 0.0cm
\includegraphics[width=8.88cm,angle=270]{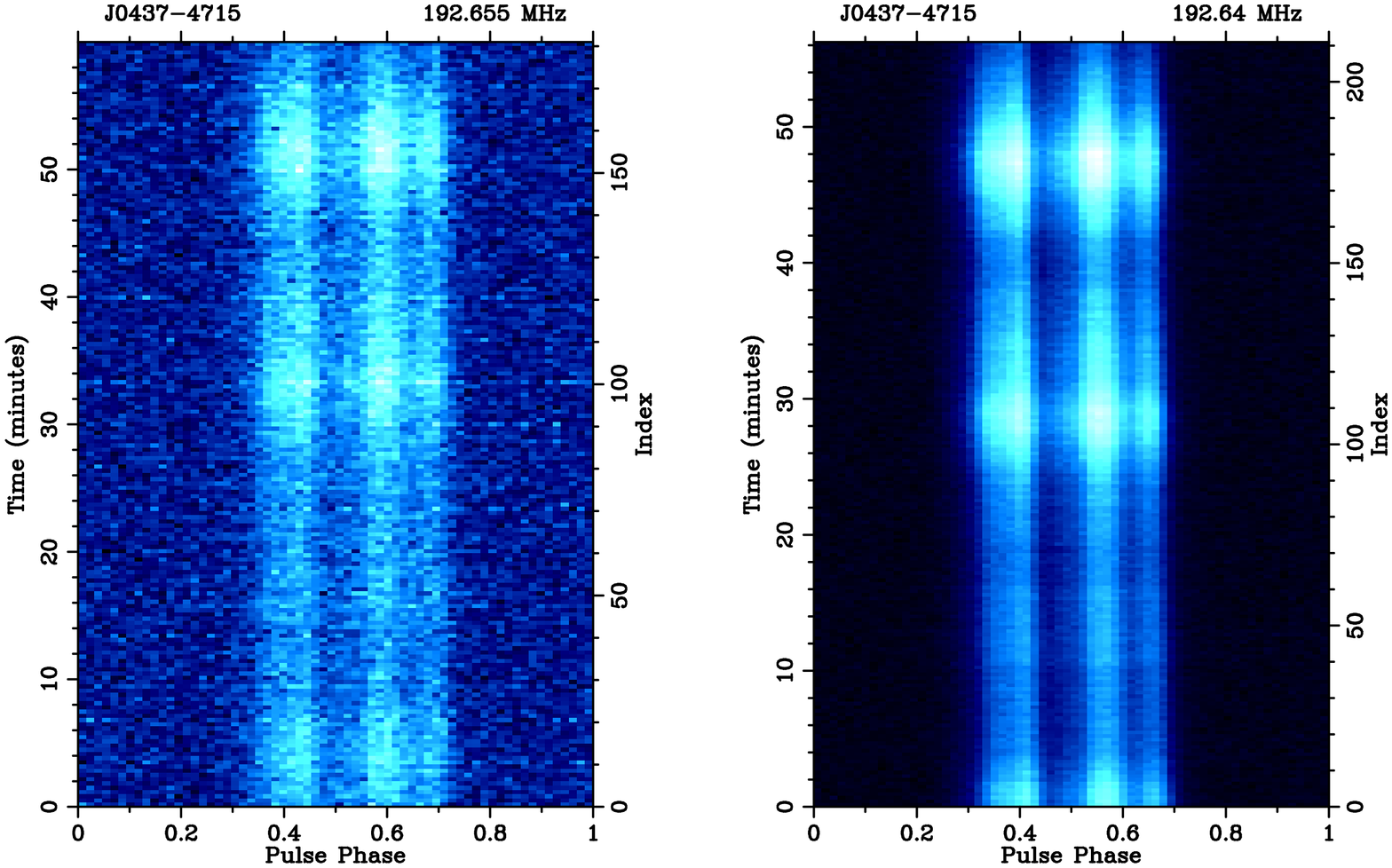}
\end{center}
\caption{MWA observations of \psr\ at a frequency of 192 MHz over a time duration of 1 hr and a bandwidth 
of 15.36 MHz.  
{\it Left}: incoherent addition of detected powers from all 128 tiles; {\it Right}: after re-processing with the 
tied-array beam-former pipeline. {\it Top}: integrated pulse profiles; {\it bottom}: pulse strength vs. time and 
pulse phase. The pulse phase resolution is approximately 90 $\mu$s. The improvement in signal-to-noise is 
over an order of magnitude. 
}
\label{fig:data}
\end{figure*}

\section{Observational data} \label{s:obs}

The MWA data used in this paper were recorded with the voltage capture system (VCS) developed for the MWA. The VCS functionality allows recording up to 24$\times$1.28 MHz from all 128 tiles (both polarisations) at the native 100-$\mu$s, 10-kHz resolutions for up to $\sim$1.5 hr \citep{tremblay+2015}, and is the primary observing mode for observations of pulsars and fast transients. Only half the recording capability (12$\times$1.28 MHz) was available in the early days of VCS commissioning when our observations were performed (MJD=56559). Further details are described in \citet{bhat+2014}. These data have now been reprocessed using our new beamformer pipeline that coherently combines voltage signals from all 128 tiles. The Parkes data used are from observations made at a frequency of 732 MHz, at an epoch two weeks later (MJD=56573) than our MWA observations.

\subsection{Tied-array beam processing of MWA observations} \label{s:tied}

A tied-array beam is  a coherent sum of voltage signals from individual tiles and is, theoretically, expected to yield a sensitivity improvement of $\sqrt{N_{\rm tile}}$ over the incoherent addition of detected powers, where \Ntile\ is the 
number of tiles. For the MWA, this means potentially an order of magnitude boost in sensitivity, besides enabling high-time resolution polarimetric studies. It involves incorporating delay models to account for the geometric and cable lengths, as well as calibration for complex gains (amplitude and phase) of individual tiles, and proper  accounting for the beam models that incorporate polarimetric response of individual tiles. The calibration and beam information are provided by an offline version of the real-time calibration and imaging system, RTS (Mitchell et al. in prep), which uses the visibilities generated from an off-line version of the MWA correlator \citep{ord+2015}. Calibration was performed using Pictor A, a bright in-beam source at $\sim$10$^{\circ}$ offset from \psr. The full processing pipeline runs on the Galaxy cluster of the Pawsey supercomputing facility\footnote{www.pawsey.org.au} that also hosts the archival VCS data after transport from the MRO. Further details on implementation of this processing pipeline are described in Ord  et al. (in prep), where we also present the first pulsar polarimetric observations with the MWA. \\

Fig.~\ref{fig:data} shows the improvement in signal-to-noise from this tied-array beam processing for \psr\ observations. The signal-to-noise ratio (S/N) of the integrated pulse profile has increased from $\sim$205 (incoherent detection) to $\sim$2100, i.e. 
an improvement  of a factor of 10, which is only $\sim$10\% less than the theoretical expectation for a coherent sum from 126 tiles\footnote{Two tiles were excluded from tied-array beam-forming owing to poor calibration solutions.}. This translates to a mean S/N$\sim$3 for individual pulses, however much larger values (S/N$\ga$10) can be expected during the times of scintillation brightening.\\

\subsection{High-resolution dynamic spectra} \label{s:high}

\begin{figure*}[t]
\epsscale{1.23}
\begin{center}
\includegraphics[width=10.10cm,angle=270]{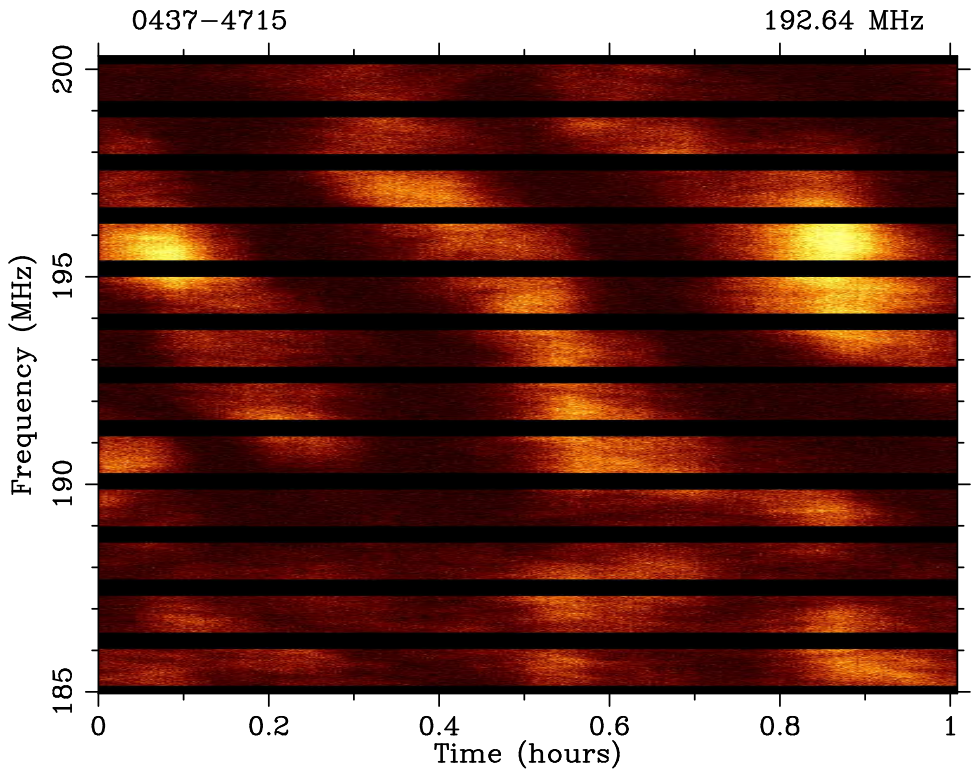}
\end{center}
\caption{Dynamic spectrum of \psr\ from MWA observations over a time duration of 1 hr 
and over a bandwidth of 15.36 MHz centred at a frequency of 192.64 MHz. 
The gaps in frequency correspond to the edge channels (20 each) on either end of a given coarse 
(1.28 MHz) channel that were not recorded due to a limitation in data recording in the early days 
of VCS commissioning. 
The data resolutions are 10 s in time and 10 kHz in frequency.   
}
\label{fig:dynsp}
\end{figure*}

The beam-formed data are processed using the {\sc dspsr} software package \citep{dspsr2011} to generate synchronously-folded pulse profiles over 10-second sub-integrations. With the  sensitivity improvement provided by the tied-array beam processing, we were able to generate a dynamic spectrum at the native frequency resolution of 10 kHz for VCS -- this is a dramatic improvement over our earlier analysis that used a spectral resolution of 640 kHz \citep{bhat+2014}. The resultant dynamic spectrum is shown in Fig.~\ref{fig:dynsp}. It is dominated by a small number of bright {\it scintles} whose intensity maxima drift in the time-frequency plane -- a consequence of refraction through the ISM. The increased sensitivity and spectral resolution results in higher sensitivity to subtle features, such as those caused by a combination of diffractive and refractive scattering effects. \\

Scintillation parameters including the characteristic scales in time and frequency (i.e. the scintillation bandwidth \nud and the diffractive time scale $\taudiff$), can be obtained from a two-dimensional auto-correlation function analysis of dynamic spectrum. The results from such an analysis, along with a detailed comparison with the published measurements, is presented in our earlier paper \citep{bhat+2014}. For the data in Fig.~\ref{fig:dynsp}, we obtain $\nud\,\sim$\,1.7\,MHz, $\taudiff\,\sim$\,260\,s, and a drift rate (in the time-frequency plane) $\dtnu\,\sim$95\,\driftunit (with measurement uncertainties $\sim$25\%).  Our measured \nud\ is discrepant with the majority of the published values, however it agrees with the larger scale of scintillation from \citet{gwinn2006}. Considering that all published measurements are from observations made at higher observing frequencies ($\sim$300-600 MHz), it is possible many of them were underestimated, particularly when the observing bandwidth ($B$) is not large enough to allow reliable measurements (e.g., $\nud \ga B$). 

\section{Secondary spectral analysis} \label{s:ss}

The dynamic spectrum is a record of the pulse intensity as a function of time and frequency, $S_1(\nu,\,t)$, and is the primary observable for scintillation analysis. Its two-dimensional power spectrum  is the secondary spectrum, 
$S_2(\fnu,\,\ft)\,=\,| S_1^{\dagger}(\nu,\, t)|^2$ (where $\dagger$ indicates two-dimensional Fourier transform). 
It is a powerful technique that captures interference patterns produced by different points in the image plane \citep[e.g.][]{cordes+2006,stine+2001}. The scatter-broadened pulsar image is seen over a field of view $\sim\Ds\thetadiff $, where \Ds is the effective distance to the screen  and \thetadiff is the half-width angular size of the broadened pulsar image. If  \vecthetaone\ and  \vecthetatwo\ are two arbitrary points in the image 
plane, the corresponding ``fringe rates"  in time and frequency, \ft\ and \fnu,  are given by, 
\be
\ft =   { (\thetatwo - \thetaone ) . \vecVeff  \over s \,\lambda }, \quad 
\fnu = { D (1-s) ( \theta _2 ^2 - \theta _1 ^2)  \over 2\,s\,c }
\ee
where $c$ is the speed of light, $\lambda$ is the observing wavelength, and $s$ is the fractional distance of the screen from the source; \fnu is a measure of the differential time delay between pairs of rays and \ft is the temporal fringe frequency. Interference between the origin and pairs along an axis in the direction of net velocity vector  \Veff\ produces parabolic scintillation arcs, represented by $\fnu=\eta\ftsq$. In essence, parabolic arcs can be described as a natural consequence of small-angle forward scattering. 

\subsection{MWA observations at 192 MHz} \label{s:intro} 

Figure~\ref{fig:arcs} shows the secondary spectrum from MWA observations at 192 MHz. 
There is a clear, albeit faint, arc-like feature in the data, particularly on the left side of the zero Doppler frequency ($\ft=0$) axis. The feature is relatively more prominent in the lower one-third of the frequency band (170--175.12 MHz), however it is still visible, with somewhat reduced strength, in data over a larger, or the full frequency range. 
Even though the full secondary spectrum  spans  fringe rates out to 50 mHz and 50 $\mu$s  for our dynamic spectral resolutions of 
$\Delta\nu$=10\,kHz and $\Delta t$=10\,s in Fig.~\ref{fig:dynsp},  the arc feature visible is largely restricted to a small region 
($\la$5\%) near the origin ($\ft\la$ 5 mHz;  \fnu\la15$\mu$s). \\

\subsection{Parkes observations at 732 MHz} \label{s:intro}

\begin{figure*}[t]
\begin{center}
\epsscale{0.5}
\includegraphics[width=5.25cm,angle=270]{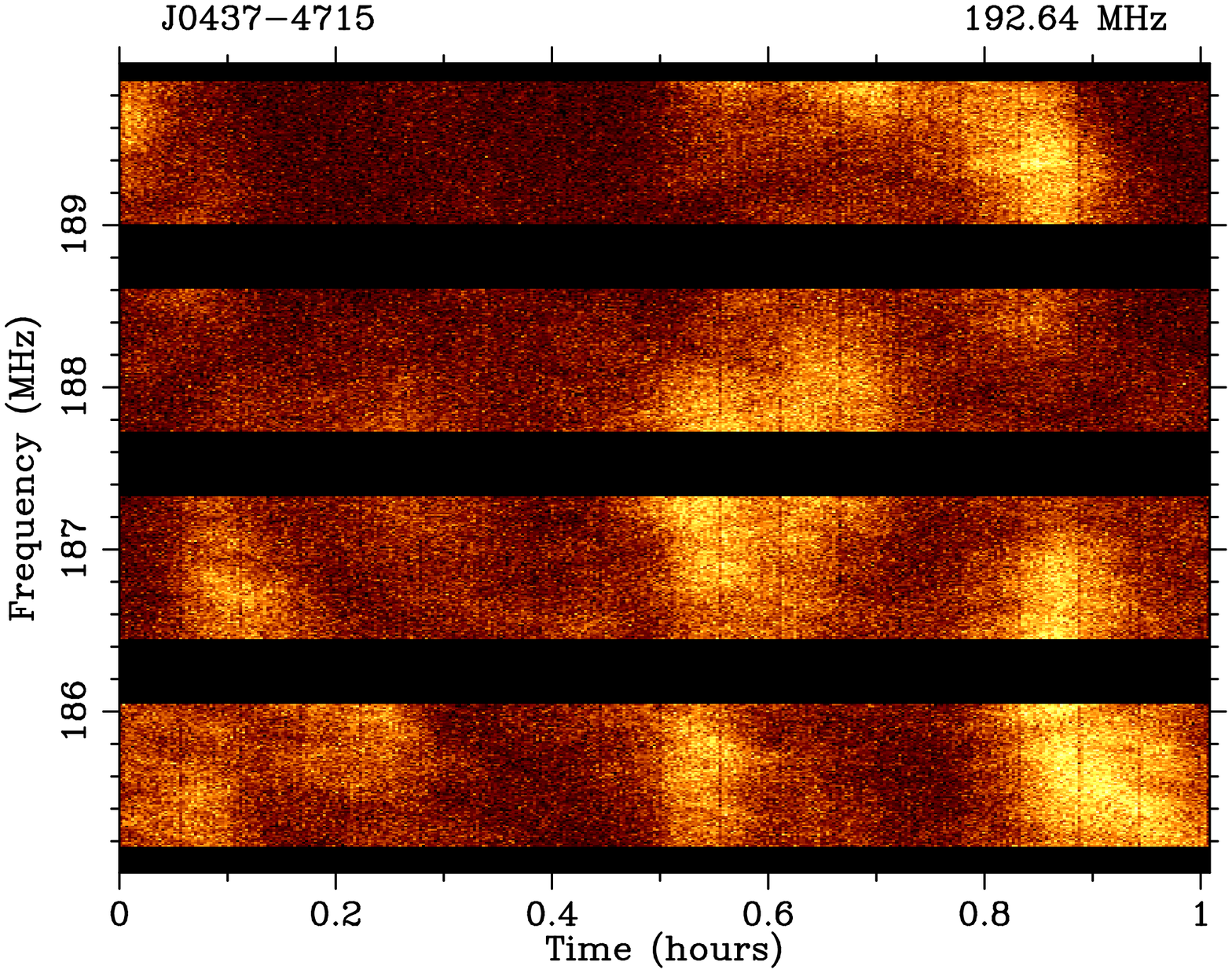}
\includegraphics[width=5.25cm,angle=270]{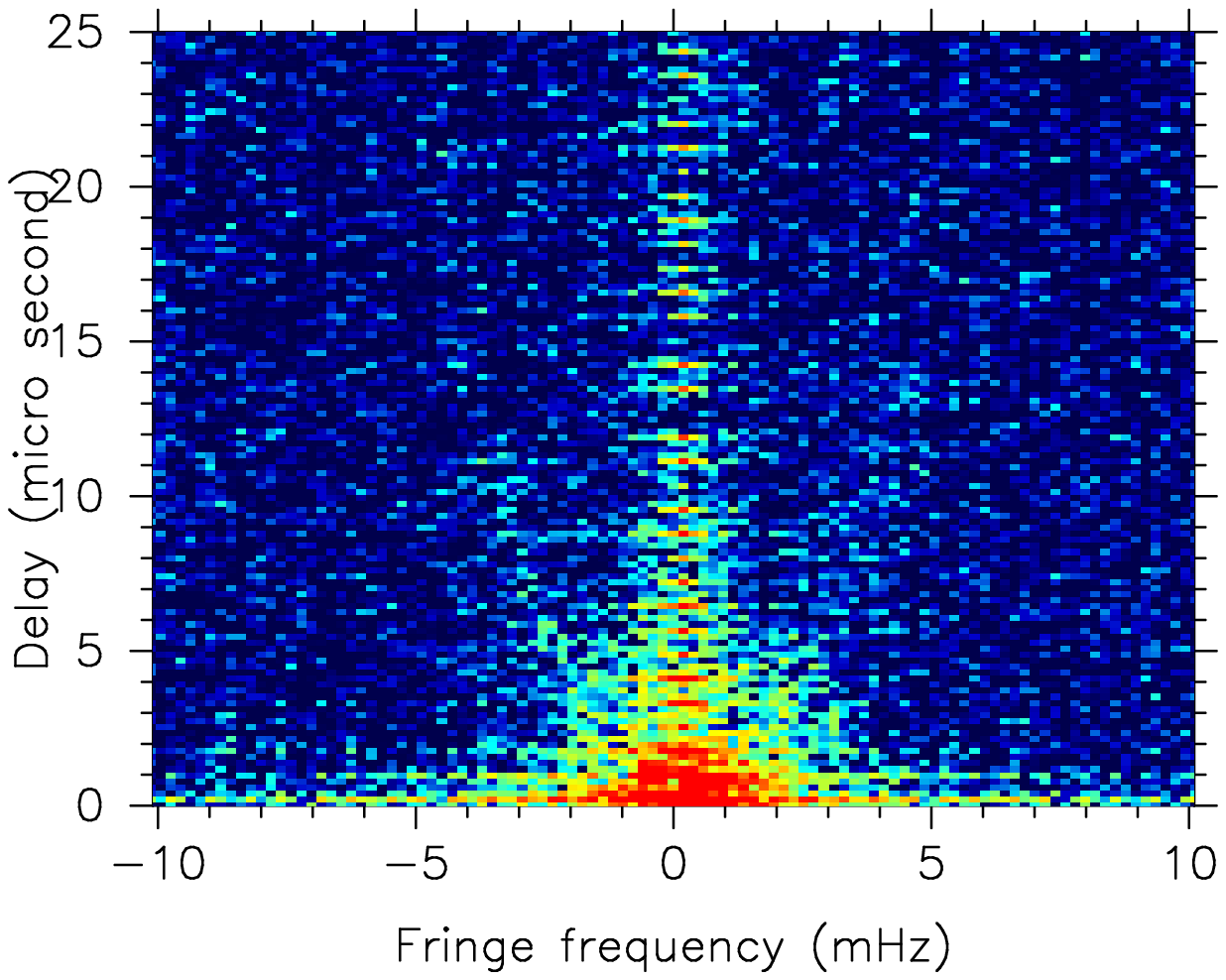}
\includegraphics[width=5.25cm,angle=270]{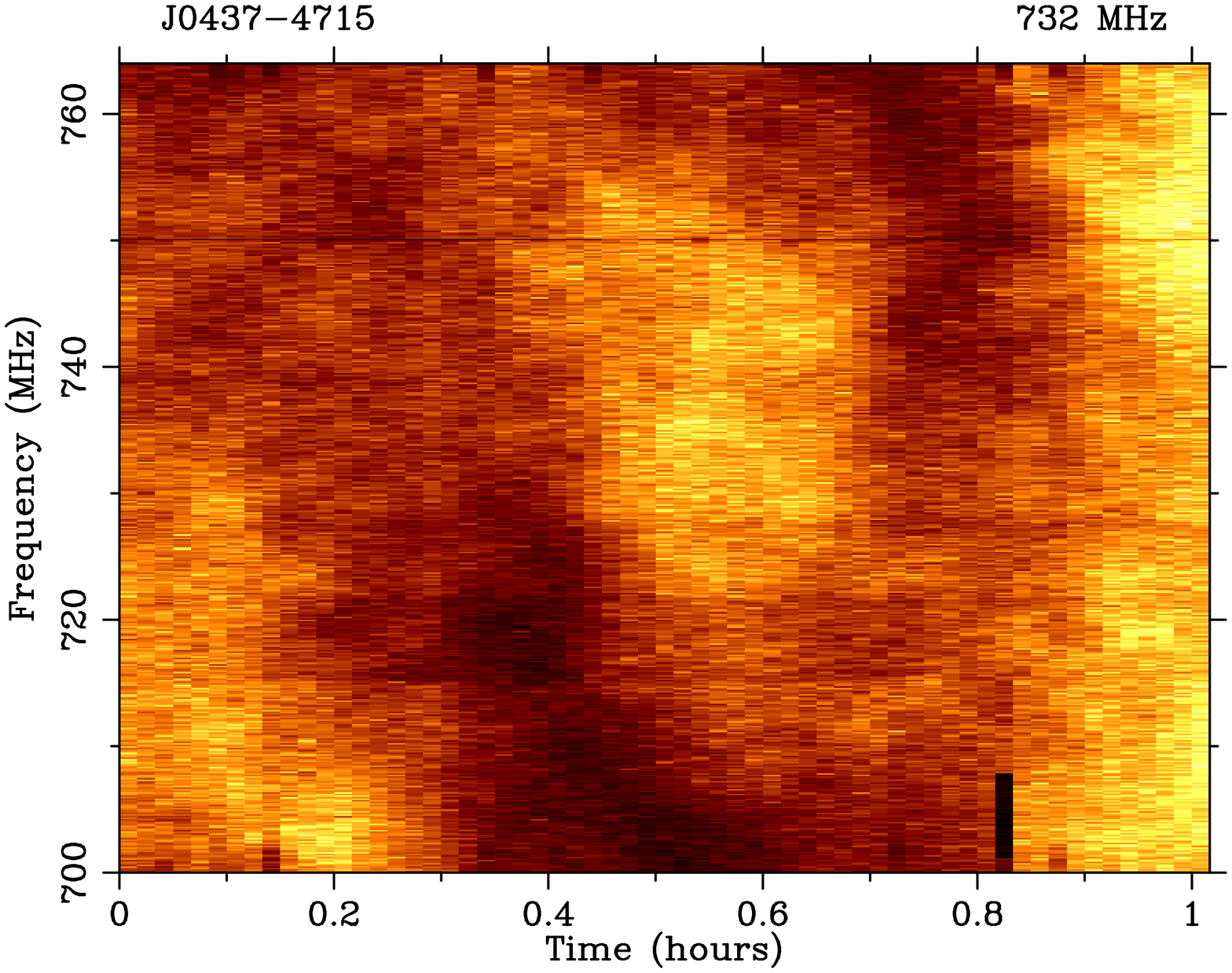}
\includegraphics[width=5.25cm,angle=270]{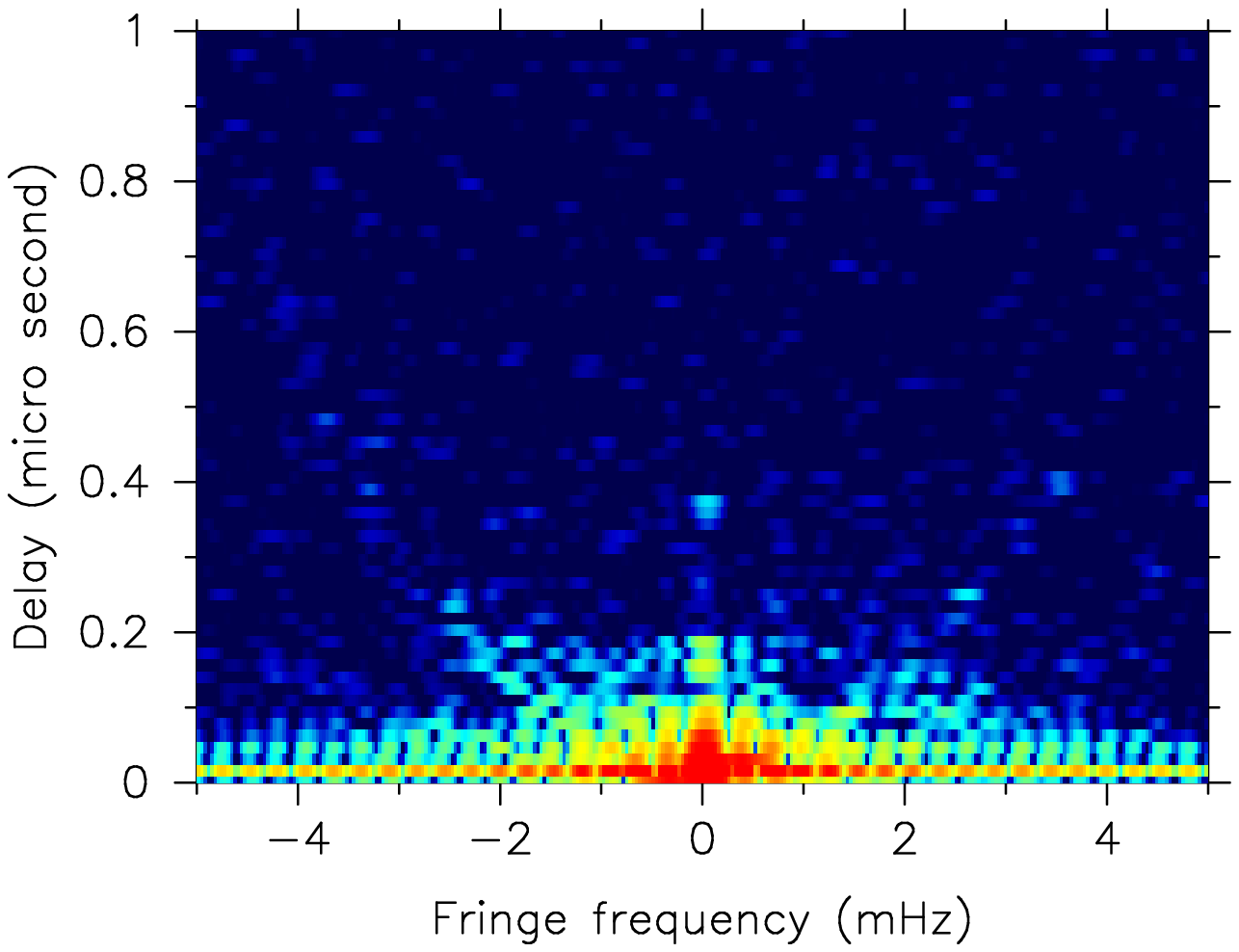}
\end{center}
\caption{ 
Dynamic spectra of \psr\ from MWA and Parkes observations (left panels) and their secondary spectra (right panels); 
MWA observations span 4$\times$1.28\,MHz near the low end of the 15.36\,MHz observing band (i.e. 170-175.12\,MHz), 
whereas Parkes observations are over a 64 MHz bandwidth centred at 732 MHz. The  resolutions in time and frequency
are 10\,s and 10\,kHz respectively for the MWA data; and 60\,s  and 125\,kHz respectively for Parkes data. For secondary 
spectra, the resolutions in the conjugate time (Doppler frequency) and conjugate frequency (delay) axes are 0.195\,mHz 
and 0.195\,$\mu$s respectively for MWA data, and 0.032\,mHz and 0.016\,$\mu$s respectively for Parkes data. 
}
\label{fig:arcs}
\vskip 0.25cm
\end{figure*}

To confirm the parabolic arc  seen in MWA data, we analysed archival Parkes data from observations at 732 MHz, the only PPTA observing frequency that is below the expected transition frequency ($\sim$1 GHz) for this pulsar. Fig.~\ref{fig:arcs} shows the dynamic spectrum from observations made at an epoch two weeks later than MWA observations.  The data were recorded with the ATNF Parkes Digital Filterbank (DFB4), and pre-processed over 1-minute sub-integrations and have a spectral resolution of 125 kHz. A parabolic arc feature is clearly visible in the secondary spectrum of these data (Fig.~\ref{fig:arcs}). There are also hints of a ``filled parabola," as seen in some of the published data at lower frequencies \citep[e.g.][]{stinebring2007}. \\

\section{Arc curvature and the placement of scattering screen} \label{s:arcs}

Theoretical treatments on parabolic arcs and the scattering geometry are discussed by \citet{stine+2001} and \citet{cordes+2006} \citep[see also][]{brisken+2010}. The fringe frequency and delay parameters \fnu\ and \ft\ can be related to the curvature of the arc ($\eta$), the pulsar distance ($D$) and proper motion ($\Vmu$), and the placement of scattering screen ($s$). As discussed by \citet{cordes+2006},  this relation depends on the number of arcs seen in observations and the scattering geometry, and in general,  $s$ can be determined to within a pair of solutions that is symmetric about $s=1/2$. 
For a screen located at a fractional distance $s$ from the pulsar, the curvature parameter $\eta$ is given by
\be
\eta =  { D\,s\,(1-s)\,\lambda^2 \over 2\,c\,\Veffsq \, {\rm cos}^2 {\alpha} }  
\label{eq:eta}
\ee
where  $ \Ds = D\,s\,(1-s) $ is the effective distance to the screen and $\alpha$ is the angle between the net velocity vector \vecVeff\ and the orientation of the scattered image. The effective velocity \Veff\ is the velocity of the point in the screen intersected by a straight line from the pulsar to the observer, which is the weighted sum of the pulsar's binary and proper motions, and the motion of the screen and the observer (\Vscreen\ and \Vobs\ respectively). Its transverse component is given by
\be
\vecVeffperp=(1-s) ( \vecVpsrperp + \vecVbinperp )  + s \, \vecVobsperp - \vecVscreenperp\
\label{eq:veff}
\ee
where \Vpsrperp\ is the transverse pulsar motion (i.e. proper motion) and \Vbinperp\ is the pulsar's binary orbital motion (transverse component). 
Thus, the measurement of $\eta$ can be used to determine the location of the scatterer, when all the contributing terms of Eq.~\ref{eq:veff} (and hence the net  \Veffperp) are precisely known. \\

The astrometric and binary orbital parameters are very well determined for \psr. Specifically, both the distance ($D$) and the proper motion ($\Vmu$) are known at very high precisions from timing and interferometric observations \citep{verbiest+2008,deller+2008}; a parallax measurement of $\pi=6.396\pm0.054$ yields $D=156.3\pm1.3$\,pc, which, when combined with
the  proper motion measurement of $\mu=141.3\pm0.1$ \muu\, implies a transverse space motion $\Vmu=105\pm1$\,\velu. The three-dimensional sky geometry of the pulsar's binary orbit is also well determined \citep{willem2001,verbiest+2008}, including the longitude of the ascending node $\Omega$. The screen velocity (\Vscreen) and the orientation angle ($\alpha$) are generally unknown; ignoring these terms (i.e. assuming \Vscreen\ is small compared to all other terms in Eq.~\ref{eq:veff} and $\alpha = 0$), leads to a simplified form for Eq.~\ref{eq:eta}, with the screen location $s$ being the sole unknown.



\begin{figure*}[t]
\begin{center}
\epsscale{0.8}
\includegraphics[width=5.25cm,angle=270]{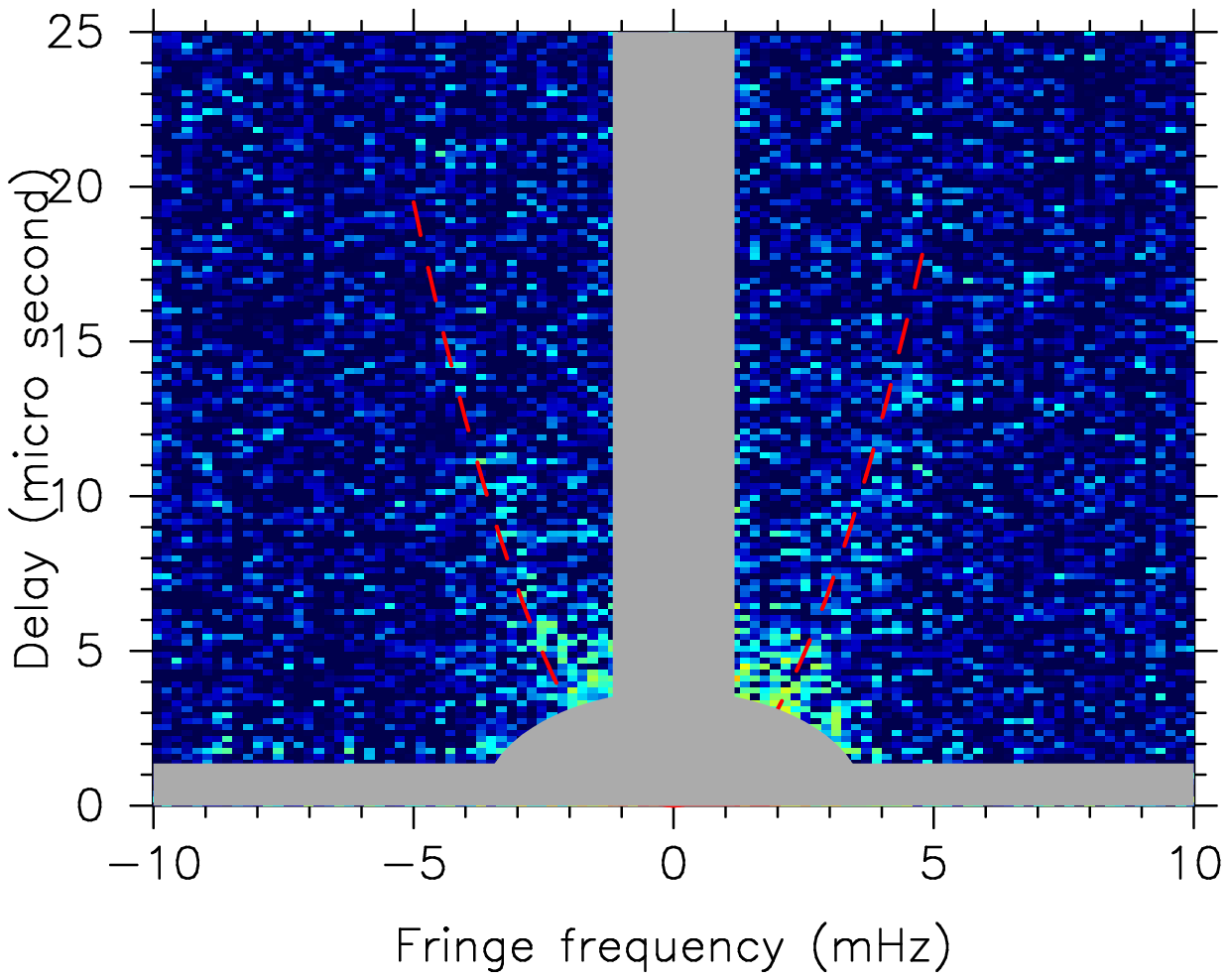}
\includegraphics[width=5.25cm,angle=270]{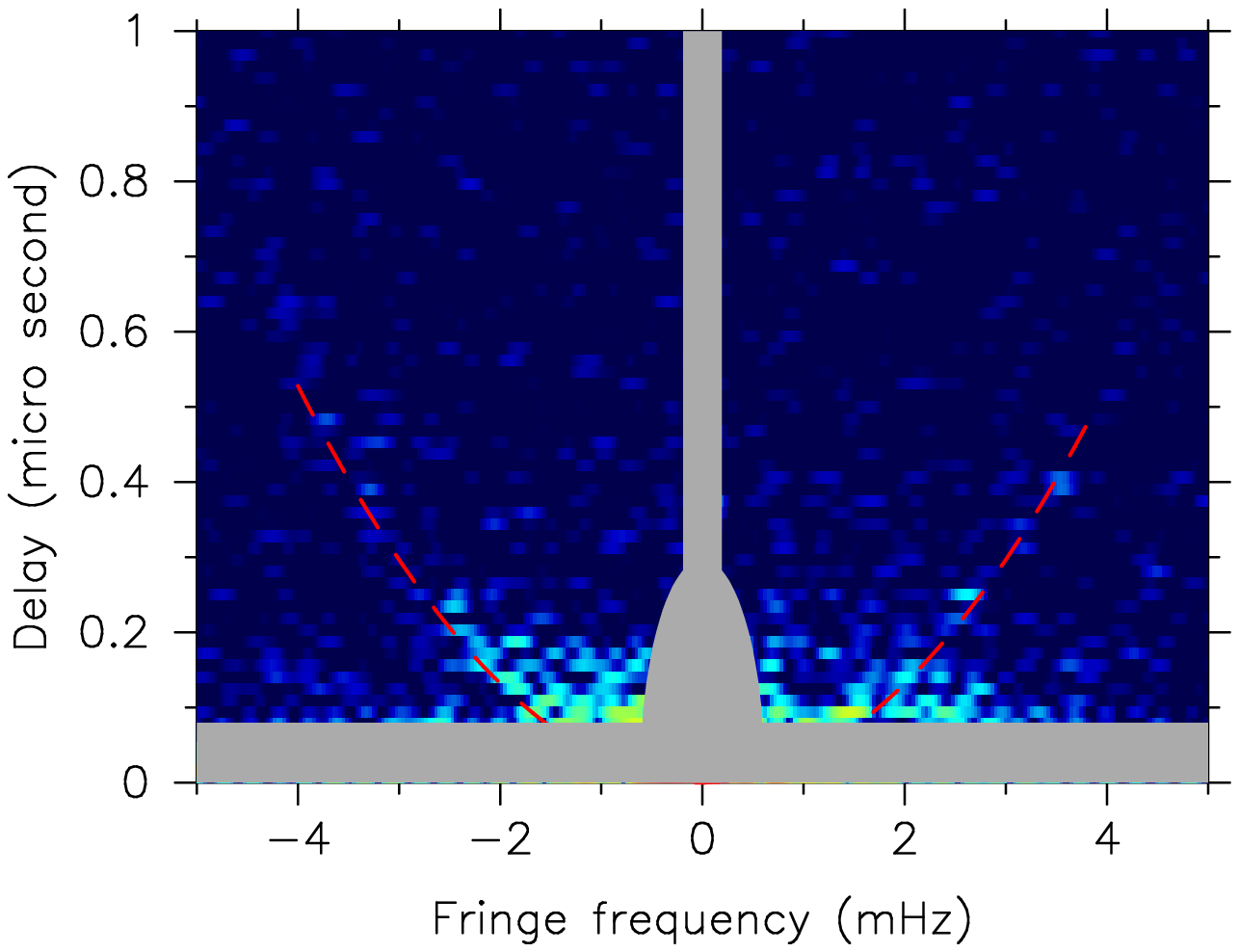}
\vskip 0.00cm
\hskip -0.25cm
\includegraphics[width=7.20cm,angle=0]{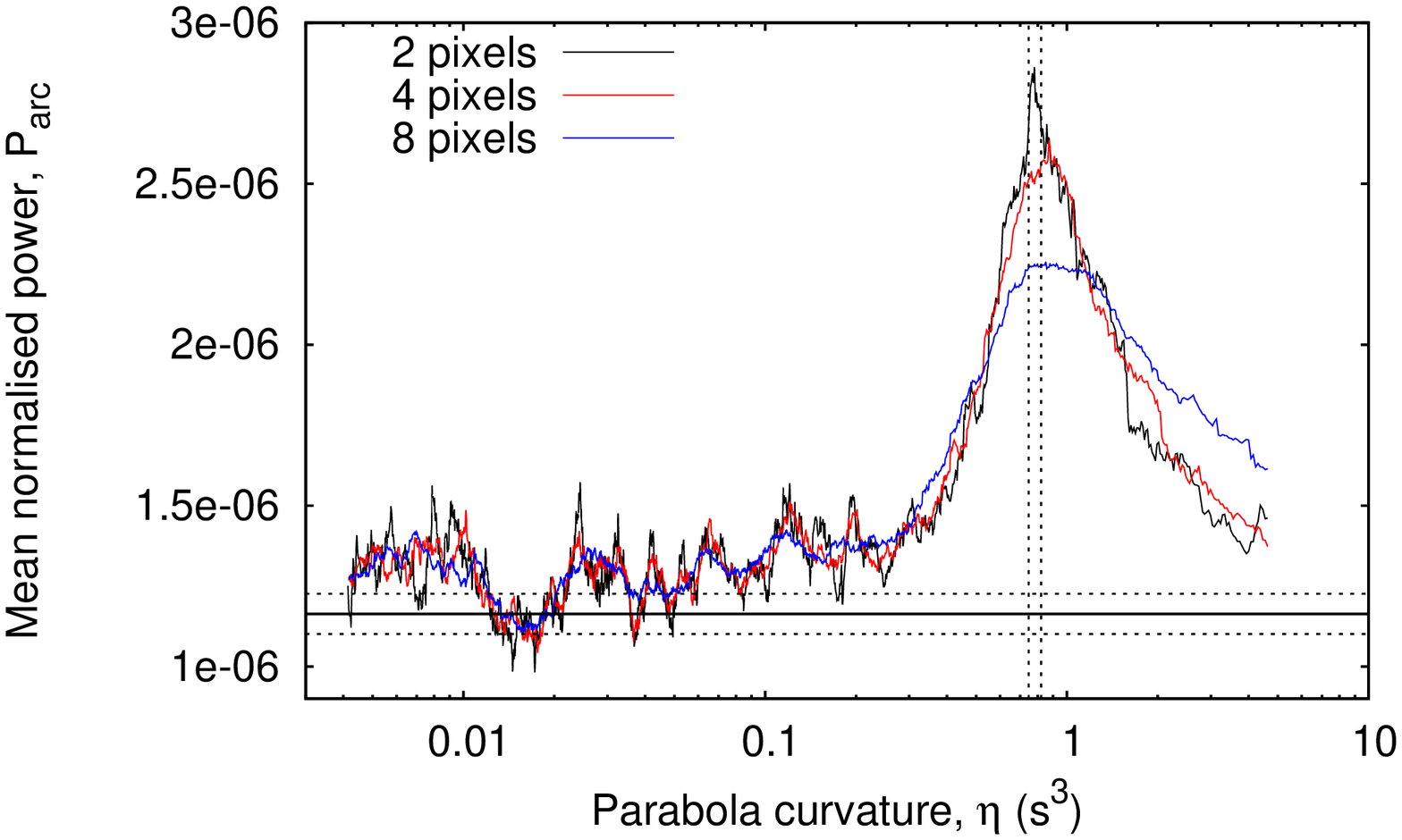}
\includegraphics[width=7.20cm,angle=0]{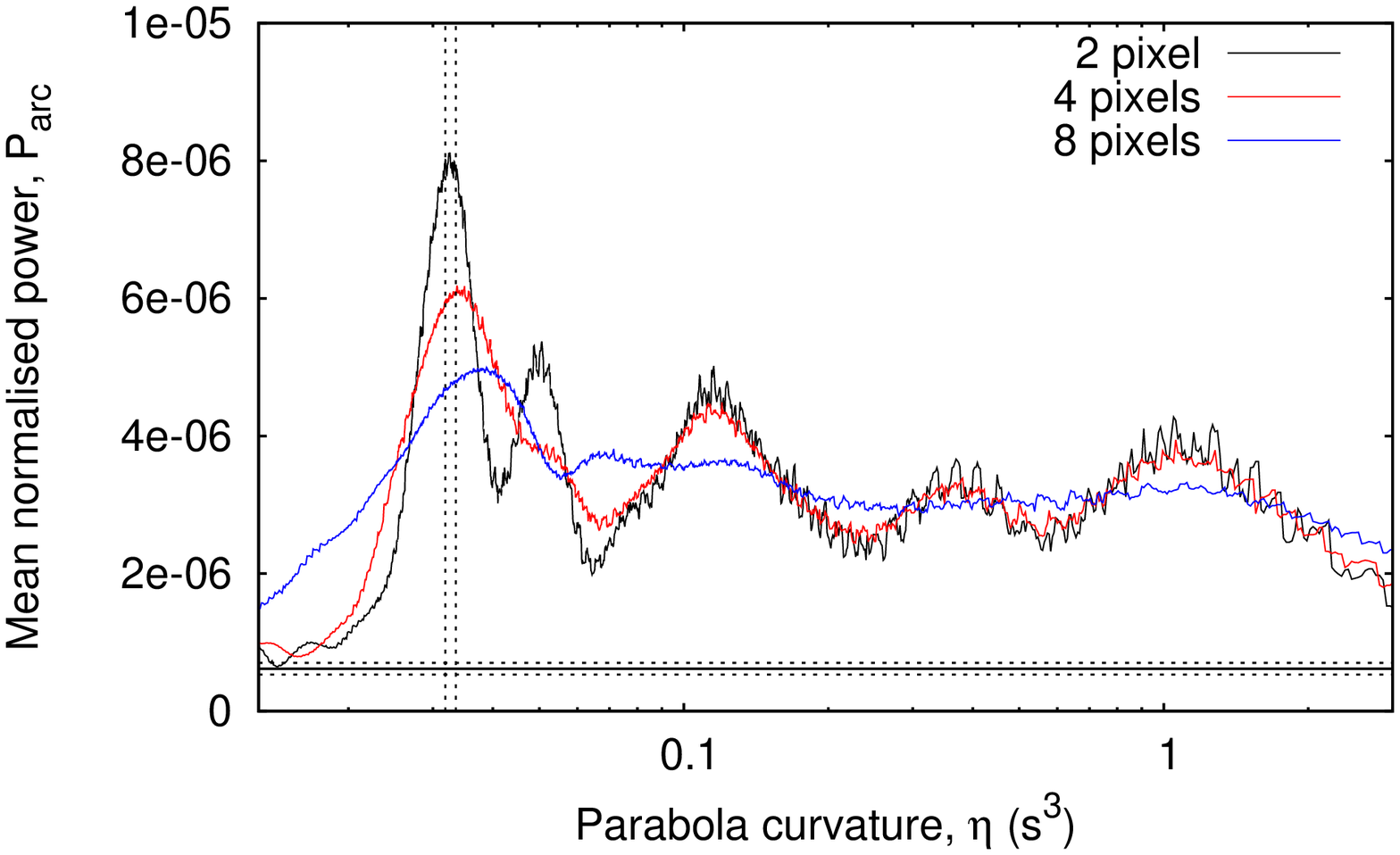}
\end{center}
\caption{
{\it Top panels}: Secondary spectra of \psr\ for MWA and Parkes data (Fig.~\ref{fig:arcs}); the shaded areas (in grey) were excluded from the analysis while estimating the arc curvature, shown as the dashed (red) curve.  {\it Bottom panels}: Mean arc strength $P _{\rm arc}$ against the curvature parameter $\eta$ for MWA and Parkes observations  (left and right panels respectively), computed outside the excluded regions of low-frequency noise (shown in grey in top panels). The 2, 4 and 8 pixels of the curves (in black, red and blue colors, respectively) correspond to the size of the `smoothing window' (i.e. thickness of the parabola) used in the computation of $P_{\rm arc}$. The solid and dotted horizontal lines correspond to the noise statistics estimated for a segment that is well outside the visible arc feature (see the text for details), whereas the dotted vertical lines correspond to the points where $P_{\rm arc}$ is 1$\sigma$ below the peak (for the two-pixel curve).
}
\label{fig:sammy}
\vskip 0.5cm
\end{figure*}

\subsection{Estimation of the arc curvature}

When observations are not limited by signal-to-noise (e.g. Arecibo data) and parabolic arcs are sharp and clearly visible, the curvature 
$\eta$ can be reliably estimated even as a best fit-by-eye \citep{stine+2001,hill+2005}. 
As the signal-to-noise of our detections are comparatively lower, we adopt a new technique that is more systematic and robust. It employs feature extraction via the application of the (one-dimensional) generalized Hough transform. 
We essentially parameterise the parabolic feature by the curvature parameter $\eta$, and trial over a wide range of values within the constraints allowed by the data, each time summing the power along the parabolic segments outside the regions of low-frequency noise 
in order to minimise contamination from high spectral values near the origin and the zero axes. The mean power computed in this manner, \Parc (henceforth refered to as ``arc strength"), is given by 
\be
P _{\rm arc} ( \eta )  = { 1 \over N } \sum _{i=1} ^N S _{2} ( \eta \ftisq, \, \fti ) 
\ee
where the summing procedure  is performed along the points of arc outside the excluded low-frequency noise, and out to delays beyond which little power is detectable (i.e. $i$=1...$N$, where $N$ corresponds to \fnu=20 $\mu$s for MWA data). \\

Fig.~\ref{fig:sammy} shows a plot of the arc strength \Parc\ against $\eta$ from our analysis, where the point of maximum mean power corresponds to the best-fit curvature.\footnote{The computation of \Parc\ is restricted to $\ft<0$ where the arc feature is more prominent.} We estimate $\eta=0.78\pm0.02$\,\etau\ for our MWA data (henceforth \etamwa) using this method. A similar analysis on Parkes data yields $\eta=0.033\pm0.001$\,\etau (henceforth \etapks), where we also note multiple secondary peaks at larger values of $\eta$ (Fig.~\ref{fig:sammy}), with a tendency for \Parc\ to plateau $\sim$2\,dB below the peak value. This is presumably arising from a sparse distribution of power inside the parabolic arc feature, suggesting scattered radiation arriving over a wider range of deflecting angles (and time delays).  A closer examination of MWA data reveals hints of a similar case, albeit comparatively weaker, but seen as a slower tailing-off of \Parc\ at larger values of $\eta$. \\

The measured values of  \etamwa\ and \etapks\  do not scale as per the theoretically-expected $\eta\propto\lambda^2$ relation. The implied scaling index of $\beta=-2.35\pm0.07$ (where $\eta\propto\lambda^{\beta}$) is  larger than the values ($-1.8\pm0.1$ to $-2.1\pm0.1$) that were   determined by \citet{hill+2003} from their observational data spanning a large frequency range (0.4 to 2.2 GHz). As we explain below, this departure from the $\lambda^2$ scaling is due to the change in \Veffperp between the two observing epochs. 

\subsection{Placement of the scattering screen} \label{s:place}

In order to determine the location of the scatterer ($s$) from the measurements of  $\eta$, we need to compute the effective velocity \Veffperp\ at the two observing epochs (cf. Eq.~\ref{eq:eta}~and~\ref{eq:veff}). As the screen velocity \Vscreen\ and the angle
$\alpha$ are generally unknown, we assume $\Vscreen \sim 0$ and $\alpha \sim 0$ in our analysis. For the observing epochs of MWA and Parkes data (MJD=56559.878 and MJD=56573.837, respectively), the corresponding true anomalies are $195.44^{\circ}$ and $350.76^{\circ}$, respectively,  i.e. a  difference in the orbital phase of 0.43 cycles. The transverse pulsar and binary motions, when projected on to the plane of the sky (with the x axis defined along the line of nodes, positive toward the ascending node), are tabulated in Table 1. 
There is a substantial change in  \Veffperp between the two epochs due the pulsar's binary motion alone; accounting for just this term, i.e. 
$\vecVeffperp \approx (1-s) (\vecVpsrperp + \vecVbinperp)$, we obtain the fractional distance from the pulsar, $s=0.27\pm0.01$, from MWA measurements and $s=0.26\pm0.01$ from Parkes measurements. \\

The contribution from the Earth's orbital motion around the Sun (\Vobsperp) can also be significant depending on the pulsar's line of sight and the observing epoch, although it will be weighted down for a screen that is located closer to the pulsar (Eq.~\ref{eq:veff}). 
The x and y components of \Vearth in the coordinate system that we have employed are tabulated in Table 1. 
The change in \Vobs between the two observing epochs is  relatively small (\la 5 \velu) in comparison to that from the pulsar's binary motion. Accounting also for this term involves solving for $s$ in Eq.~\ref{eq:eta}, which yields a pair of solutions, $s=0.26\pm0.01$ and $s=0.97\pm0.01$ from \etamwa, and $s=0.27\pm0.01$ and $s=0.98\pm0.01$ for \etapks. However, in light of our previous, independent analysis based on scintillation measurements \citep{bhat+2014}, which yields a scintillation velocity \Viss = $325\pm90$ \velu $\sim 3 \,\Vmu $, i.e. \Viss $>$ \Vmu, and therefore the solution suggesting a screen closer to the pulsar is favoured.\footnote{$s$=0.97 would imply \Viss $\sim$ 0.03 \Vmu, which is not supported by observations.}
The implied screen placements are $115\pm2$\,pc and $114\pm2$\,pc  respectively from \etamwa\ and \etapks, or effectively $115\pm3$\,pc (from the Earth) if we combine the two estimates. 

\begin{deluxetable*}{ccccccc}[t]
\tablewidth{0pc}
\tablecaption{Transverse pulsar and earth motions$^a$ and the estimated screen placements \label{tab:scnt}}
\tablehead{
\colhead{Observation} & \colhead{Frequency} & \colhead{\Vpsrperp} &   \colhead{\Vbinperp} &  \colhead{\Vobsperp} &  \colhead{Arc curvature, $\eta$} &  
\colhead{Screen location ($s$)} \\
 & (MHz) &  (\velu) & (\velu) & (\velu) & (\etau) &
}
\startdata
MWA  & 192 & ($-55.2, -89.3$) & ($18.9, 1.1$)    & ($-21.97, 18.19$) & $0.78\pm0.02$     & $0.26\pm0.01$ \\
Parkes & 732 & ($-55.2, -89.3$) & ($-16.5, -6.8$) & ($-26.18, 12.81$) & $0.033\pm0.001$ & $0.27\pm0.01$ 
\enddata
\tablecomments{$^a$The x and y components when projected on to the plane of sky, with the x axis along the line of nodes (see the text for details).}
\end{deluxetable*}

\medskip

\section{Discussion} \label{s:disc}

Our results in terms of the screen placements derived from MWA and Parkes observations are in very good agreement, despite the fact that the observations were made at widely separated observing frequencies and not contemporaneous. The time separation of two weeks is significantly longer than the expected refractive time scale ($\tauref$), i.e. the characteristic time on which a new volume of scattering material is expected to move across the pulsar's line of sight. It is given by $\tauref \sim ( 2 \nuobs / \nud) \, \taudiff$, where $\nuobs$ is the frequency of observation, $\nud$ and $\taudiff$ are the scintillation bandwidth and   diffractive time scale, respectively, both of which are measurable from dynamic spectra. For our MWA observations,  $\nud \sim$ 1.7 MHz and $\taudiff \sim$ 4.5 minutes \citep{bhat+2014}, and hence $\tauref \sim$\,17 hr~\la~1 day. This estimate  is however not so reliable since it is based on single-epoch measurements, but even then it is unlikely \tauref\ may be longer than $\sim$ a few days at the MWA's frequency.  Our observational results may therefore suggest that the underlying scattering structure persists over multiple refractive cycles. The existence of  such large scattering structures in the ISM was also suggested by past observations including those where the drift slopes and multiple imaging episodes were seen to persist over time scales of $\sim$ several months \citep[e.g.][]{gupta+1994,rickett+1997,bhat+1999}. \\

\medskip

Another subtlety pertains to the ISM volume sampled by multipath scattering, which is a strong function of the observing frequency. As discussed in \S~\ref{s:ss}, the scatter broadened pulsar image has a characteristic size $\sim\Ds\,\thetadiff$, and consequently the ISM sampled by MWA observations is more than two orders of magnitude larger than Parkes observations (since \thetadiff $\propto$ $\lambda^2$). Although often ignored in observational interpretations, this can be potentially an important effect. Recent work of \citet{cordes+2015} explores this in great detail in the context of frequency-dependent (chromatic) DMs in timing-array observations. Nonetheless, it is not yet clear how this may influence scattering and scintillation observables and their scalings with the frequency. There is no compelling observational evidence in support of chromatic DMs, and  a wealth of observational data on scintillation and scattering measurements are seen follow the expected frequency scaling  over a large range. In particular, we note the work of \citet{hill+2003}, who experimentally verified the scaling relation for scintillation arcs; their derived scaling indices range from $-1.8\pm0.1$ to $-2.1\pm0.1$, and are consistent with the $\lambda^2$ scaling, despite observational data spanning a large frequency range (from 0.4 to 2.2 GHz).  Contemporaneous observations at multiple different frequencies, similar to those advocated by \citet{lam+2015} for improved DM corrections in PTA observations, will be useful for gaining further insights into 
this aspect. \\

Aside from these subtleties, our observations of scintillations arcs are clear indications of scattering toward \psr\ arising from a localized region (thin screen).  The implied screen location of $121\pm3$\,pc is, incidentally, consistent with the expected location of $\sim$100-120\,pc to the edge of the Local  Bubble \citep{snowden+1990,bhat+1998,ne2001,spangler2009}.
The possibility of the screen being closer to the pulsar was also hinted by our earlier, independent analysis, where our measured scintillation velocity (\Viss = $325\pm90$\,\velu) suggested a screen location of $\sim80-120$\,pc from the Earth (i.e. $s\sim0.25 - 0.5$) based on \Viss/\Vmu$\sim$3 \citep{bhat+2014}. Scattering toward \psr\ therefore most likely dominated by the material near the edge of the bubble. \\

Even as our observations of scintillation arcs suggesting a small fraction of the scattered radiation arriving at large delays, its impact on timing precision may be negligibly small for this pulsar at its timing frequencies of  $\sim$1-3\,GHz. This is because \psr\ is a weakly-scattered pulsar, with the second lowest value for the measured strength of scattering (the wavenumber spectral coefficient, $\Cnsq\sim9\times10^{-5}$ \cnsqunits from our measurement of scintillation bandwidth). Based on MWA observations, a  transition to weak scintillation can be expected near $\sim$1\,GHz, and consequently scattering effects are no longer relevant at frequencies $\ga$1\,GHz. However, this will not be the case for many other PTA pulsars. The DM range of PTA pulsars extends out to $\sim$300 \dmu, even though the majority of them are at DMs \la 50 \dmu. Since scattering delays ($\taud$) are expected to scale as ${\rm DM^{2.2}}$, timing perturbations $\sim$100 ns can be expected for PTA pulsars at the $\sim$1-2 GHz timing frequencies. DM variations may still be the dominant source of ISM noise in PTA data; however, scattering effects may also be important, particularly if DM corrections are to rely on observations at $\la$1\,GHz. Observations at the low frequencies with MWA, LWA and LOFAR, and eventually with SKA-LOW, can therefore prove to be very useful in assessing the importance of scattering delays and the nature of turbulent ISM toward PTA pulsars. \\

\section{Conclusions and future work} \label{s:conc}

A new processing pipeline for MWA high time resolution data enables forming a coherent combination of tile powers from recorded voltages, bringing an order of magnitude improvement in the sensitivity for pulsar observations. We have demonstrated one of its applications, through high-resolution dynamic spectral studies of  \psr\ from MWA observations at 192\,MHz. A secondary spectral analysis reveals parabolic scintillation arcs, whose curvature scales  as $\lambda^2$ to Parkes observations at  732 MHz, once accounted for the change in the net effective velocity due to the pulsar's binary orbital and the Earth's motions. Our analysis suggests that scattering toward \psr\ predominantly arises from a compact region located about $\sim$115\,pc from the Earth, which is comparable to the distance to the edge of the Local Bubble ($\sim$100-120\,pc) that encapsulates the local Solar neighbourhood. Dedicated observational campaigns at the low frequencies of MWA and LOFAR, preferably contemporaneously with timing-array observations made at higher frequencies, can be potentially promising for a detailed characterisation of  the ISM along the lines of sight, and for assessing the sources of ISM noise in timing-array data. \\

\noindent
{\it Acknowledgments}: 
We thank an anonymous referee for several insightful comments which helped improve the content and presentation of this paper.
We  also thank J.-P.~Macquart, R.~M.~Shannon, M.~Bailes and H.~Knight for several useful discussions. 
This scientific work makes use of the Murchison Radio-astronomy Observatory, operated by CSIRO. We acknowledge the Wajarri Yamatji people as the traditional owners of the Observatory site.   
NDRB is supported by a Curtin Research Fellowship. 
Support for the operation of the MWA is provided by the Australian Government Department of Industry and Science and Department of Education (National Collaborative Research Infrastructure Strategy: NCRIS), under a contract to Curtin University administered by Astronomy Australia Limited. We acknowledge the iVEC Petabyte Data Store and the Initiative in Innovative Computing and the CUDA Center for Excellence sponsored by NVIDIA at Harvard University, and support from the Centre for All-sky Astrophysics (CAASTRO) funded by grant CE110001020.



\bibliographystyle{apj}

\end{document}